# A study of resting-state EEG biomarkers for depression recognition

Shuting Sun, Jianxiu Li, Huayu Chen, Tao Gong, Xiaowei Li[*], Bin Hu[*]

Gansu Provincial Key Laboratory of Wearable Computing, School of Information Science and Engineering, Lanzhou University, Lanzhou, China

## Abstract

**Background:** Depression has become a major health burden worldwide, and effective detection depression is a great public-health challenge. This Electroencephalography (EEG)-based research is to explore the effective biomarkers for depression recognition.

**Methods:** Resting state EEG data was collected from 24 major depressive patients (MDD) and 29 normal controls using 128 channel HydroCel Geodesic Sensor Net (HCGSN). To better identify depression, we extracted different types of EEG features including linear features, nonlinear features and functional connectivity features phase lagging index (PLI) to comprehensively analyze the EEG signals in patients with MDD. And using different feature selection methods and classifiers to evaluate the optimal feature sets.

**Results:** Functional connectivity feature PLI is superior to the linear features and nonlinear features. And when combining all the types of features to classify MDD patients, we can obtain the highest classification accuracy 82.31% using ReliefF feature selection method and logistic regression (LR) classifier. Analyzing the distribution of optimal feature set, it was found that intrahemispheric connection edges of PLI were much more than the interhemispheric connection edges, and the intrahemispheric connection edges had a significant differences between two groups.

**Conclusion:** Functional connectivity feature PLI plays an important role in depression recognition. Especially, intrahemispheric connection edges of PLI might be an effective biomarker to identify depression. And statistic results suggested that MDD patients might exist functional dysfunction in left hemisphere.

## Introduction

Depression is a common mental disease characterized by persistent low mood, anhedonia, grief and cognitive impairment that severely affects people's quality of life. The prevalence of depression is about 2 - 4% worldwide, and 1.7 - 2% in China[1]. According to statistics of the World Health Organization (WHO), more than 350 million people suffered different degrees of depression worldwide[2]. And a related study of a meta-analysis of 50371 patients from 118 studies found that the correct recognition rate for depression was only 47.3%[3]. So with the high incidence and low recognition rate of depression, to explore simple, objective, accurate evaluation methods or biomarkers for depression detection is a major public-health challenge[4].

At present, the diagnosis method of depression depends on the physician consultation and scale assessment. There are many obvious disadvantages, such as patient denial, poor sensitivity, subjective biases and inaccuracy[5]. In recent years, many neuroimaging technologies including positron emission tomography (PET), functional magnetic resonance imaging (fMRI) and electroencephalography (EEG) have been used to study mental illness, such as schizophrenia[6] and depression[7]. Although, PET and fMRI provide a high spatial resolution, compared with EEG, these technologies are expensive and complicated to operate. And PET requires injecting radioactive substances into the subjects, while fMRI is not suitable for the subjects with claustrophobia. So EEG with its advantages of high temporal resolution, non-invasive, easy to record and relatively low cost is the optimal choice for studying depression. Especially, there was a study indicated that the cognitive ability of depressive patients changed with mood, and these changes influenced EEG[8]. Therefore, this study will explore depression based on EEG signals.

[*] Corresponding author. Email address: {lixwei, bh}@lzu.edu.cn

Traditional researches usually extract linear and nonlinear features from EEG signals to identify depression. For example, Erguzel et al. [9] extracted frequency domain linear features and used back-propagation neural network (BPNN) with genetic algorithm (GA) to classify 147 severely depressed patients with an accuracy of 89.12%. Hosseinifard et al. [10] extracted power ratings of four EEG bands and four nonlinear features including detrended fluctuation analysis, higuchi fractal, correlation dimension and lyapunov exponent to classify 45 depressed patients and 45 normal subjects. A highest classification accuracy of 90% is obtained by all four nonlinear features and logistic regression (LR) classifier. Yang et al. [11] used 10 time-frequency linear features and 8 nonlinear features for classifying 17 depressed patients and 17 normal subjects, highest classification accuracy yielded about 83.24% when using a voting strategy. And Cai et al. [12] also extracted linear features and nonlinear features and used k-nearest neighbor (KNN) with GA to classify 86 depressed patients and 92 normal controls with a classification accuracy 86.98%. However, the human brain is composed of dynamic neural communications and the interactions based on synchronous oscillation among different brain areas[13, 14]. Recently, a considerable number of literatures have studied the synchronous oscillation mode reflecting brain activity, and provided reliable markers for brain function or dysfunction[15]. Therefore, simple linear and nonlinear analysis cannot extract all the information contained in the EEG signals.

Moreover, depression is a system-level disease, which is related to the dysfunction of neuronal network activity across multiple brain areas, rather than the breakdown of a single brain area[16]. So more and more studies used EEG signals to evaluate the differences of functional connectivity between the patients with major depressive disorder (MDD) and normal controls (NC)[17], and classify them using these differences. For example, Erguzel T T et al. [18] used coherence and adopted support vector machine (SVM) with Imporved Ant Colony Optimization (IACO) to classify 46 bipolar disorder and 55 depressed patients with a classification accuracy 80.19%. W Mumtaz et al. [19] extracted synchronization likelihood and used SVM with rank-based feature selection method based on receiver operating curve to classify 34 depressed patients and 30 normal subjects with an accuracy 98.00%. L Orgo et al. [20] extracted functional connectivity feature coherence and graph theory measures including clustering coefficient, characteristic path length, small-worldness and adopted SVM with GA to classify 37 depressed patients and 37 normal controls with a classification accuracy 88.10%. Leuchter A F et al. [21] also extracted coherence to classify 121 depressed patients and 37 normal controls and obtained the classification accuracy of 81%. And Peng H et al. [22] used phase lagging index (PLI) and adopted SVM and an altered Kendall rank correlation coefficient to classify 27 depressed patients and 28 healthy volunteers with an accuracy 92%. So these researches all prove that functional connectivity as the feature has the potential to detect depression. But what should we noted is that previous studies often used coherence to construct functional connectivity, which is strongly affected by the artefacts of volume conduction[22]. So to overcome this issue, this paper will adopt the robust to artifacts of volume conduction method PLI to calculate functional connectivity based on EEG signals of MDD.

Furthermore, due to the pathophysiology and underlying of neurological mechanism of depression are still unclear, this paper will extract EEG features from multiple perspectives, including linear, nonlinear and functional connectivity features from pre-processed EEG data. To the best of our knowledge, few studies used these features together so as to achieve a comprehensive analysis of EEG signals in patients with depression. Therefore, the first purpose of this paper was to evaluate the optimal feature sets using different feature selection methods and classifiers, the second purpose was to evaluate the optimal machine learning framework of depression detection, the third purpose was to analyze the distribution characteristics of brain regions of the optimal feature sets, and further to explore the abnormal brain pattern of depression.

## Material and methods

### Participants

A total of 24 outpatients (female/male = 11/13, 30.88 ± 10.37 years old) diagnosed with depression, as well as 29 NC (female/male = 9/20, 31.45 ± 9.15 years old) were recruited for the study. There were no significant between-group differences in age (t = 0.214, p = 0.832), or gender ($\chi 2$ = 1.224, p = 0.269). Patients with MDD were recruited among inpatients and outpatients from Lanzhou University Second Hospital, Gansu, China, diagnosed and recommended by professional psychiatrists. The NC were recruited by posters. The study was approved by the Ethics Committee of the Second Affiliated Hospital of Lanzhou University, and written informed consent was obtained from all subjects before the experiment began. All MDD patients received a structured Mini-International Neuropsychiatric Interview (MINI)[23] that met the diagnostic criteria for major depression of Diagnostic and Statistical Manual of Mental Disorders (DSM) based on the DSM-IV[24].

The inclusion criteria for all participants including: (1) the age should between 18 and 55 years old; (2) primary or higher education level. The exclusion criteria for all participants including: (1) abused or dependent alcohol or psychotropic drugs in the past year; (2) women who were pregnant and in lactation or taking birth control pills. For MDD patients, the inclusion criteria were: (1) The patients with depression met the diagnostic criteria of The MINI, and their Patient Health Questionnaire-9item (PHQ-9)[25] score was ≥ 5; (2) no psychotropic drug treatment taken within two weeks. For MDD patients, the exclusion criteria were: (1) having mental disorders or brain organ damage; (2) having a serious physical illness; (3) having severe suicidal tendency. For NC, the exclusion criteria including: (1) a personal or family history of mental disorders.

Before the experiment, the self-reported PHQ-9, and Generalized Anxiety Disorder-7 (GAD-7) [26] were self-rated in all subjects. As expected, patients exhibited greater scores in PHQ-9 (MDD: 18.33 ± 3.50, NC: 2.66 ± 1.80), and GAD-7 (MDD: 13.42 ± 4.94, NC: 2.10 ± 2.08) relative to NC (p < 0.001). After completion of the experiment, each participant received a reward for participating in this experiment.

### EEG recording and processing

5 minutes eye-closed resting state EEG was recorded using a 128-channel HydroCel Geodesic Sensor Net (Electrical Geodesics Inc., Oregon Eugene, USA) and Net Station acquisition software (version 4.5.4). The sampling frequency was 250 Hz. All the raw electrode signals were referenced to the Cz. The impedance of each electrode was kept below 50 kΩ[27]. The multi-channel EEG signals were collected in a quiet, sound-proof, well-ventilated room without strong electromagnetic interference. Participants completed the tasks sitting alone in the room, while the operators were monitoring their progress in the adjoining room. In the experiment, participants were required to keep awake and still without any bodily movements including heads or legs, as well as any unnecessary eye movements, saccades, and blinks.

The raw EEG signals were further preprocessed offline with the EEGLAB toolbox [1] in MATLAB. First, the EEG recordings were filtered between 1 Hz and 40 Hz by a Hamming windowed Sinc FIR filter, which could discard the electrical interference from the 50 Hz frequency noise and the baseline drift. Second, electrooculography (EOG) and electromyography (EMG) artifacts were removed by the TrimOutlier plugin[2]. Third, bad channels and bad data points were eliminated repeatedly by the threshold value set by mean and standard deviation. Then, the position of the removed bad channel was interpolated by spherical interpolation. Forth, the processed EEG data were re-referenced against REST[28]. Fifth, after finished the above steps, the remaining data points contained some high-power content, and some EEG epochs were removed by Artifact Subspace Reconstruction (ASR) plugin[3]. In the final, to ensure that the number of epochs is same between subjects, the 40 2 s valid epochs (40 * 2 s = 80 s) without artefacts were selected from each subject for further feature extraction. It is worth noting that for reasons of time performance and computational efficiency, we chose 16 electrodes (Fp1/2, F3/4, C3/4, P3/4, O1/2, F7/8, T3/4, T5/6) as opposed to 128 electrodes. Moreover, previous studies have widely adopted these electrodes for depression research[29, 30].

**Feature extraction**

In this study, linear, nonlinear and functional connectivity features were extracted from pre-processed EEG data to achieve comprehensive analysis of EEG signals in patients with depression.

**EEG Linear feature**

The linear features include variance; mean amplitude of peak to peak; mean square; activity, mobility and complexity which were calculated based-on time-varying Hjorth[31] parameters; max power density, the peak power frequency and the power density integral which were calculated based on adaptive Auto Regressive model. In total, we got 8 linear features. So the feature vectors dimension of linear features was 128 (16 electrodes * 8 linear features).

**EEG Non-linear feature**

The nonlinear features[10, 32] include C0-complexity; Singular-value deposition entropy (SVDen); Spectral entropy; and chaotic time series analysis of Rayleigh entropy (Renyi_entropy) including three features. In total, we got 6 nonlinear features. So the feature vectors dimension of nonlinear features was 96 (16 electrodes * 6 nonlinear features).

**EEG Functional connectivity feature**

To construct functional connectivity matrix, we adopted PLI[33] to calculate functional connectivity matrices based on sensor layer EEG signals of MDD, which was robust to artifacts of volume conduction. PLI is used to estimate the asymmetry of the phase differences distribution between two-channel EEG signals, PLI is defined as:

$$\text{PLI}_{xy}(f) = |<sign(\emptyset_x(f) - \emptyset_y(f))>| \qquad (1)$$

Where <> represents expectation value, $\emptyset_x(f) - \emptyset_y(f)$ represents the phase synchronization between the signals in channel x and y at the frequency of f. It's essential to know the instantaneous phase of the two signals involved, which can be completed using the analytical signal based on the Hilbert transform[34], so as to compute the phase synchronization. Thus, for each subject, we obtained a 16 × 16 symmetric functional connectivity matrix $FC_{ij}$:

$$FC_{ij} = \begin{bmatrix} FC_{11} & FC_{12} & \cdots & FC_{1n} \\ FC_{21} & FC_{22} & \cdots & FC_{2n} \\ \vdots & \vdots & \ddots & \vdots \\ FC_{n1} & FC_{n2} & \cdots & FC_{nn} \end{bmatrix}_{16 \times 16}$$

---

[1] EEGLAB Wiki: https://sccn.ucsd.edu/wiki/EEGLAB
[2] TrimOutlier plugin: https://sccn.ucsd.edu/wiki/TrimOutlier
[3] Clean_rawdata (ASR): http://sccn.ucsd.edu/eeglab/plugins/ASR.pdfs

16 is the electrode channels, the row i and column j of matrix represents the connectivity strength of channels i and j. The value of $FC_{ij}$ is between [0, 1], where 1 indicates perfect phase synchronization, 0 indicates no coupling. In order to use functional connectivity values as classification features, we removed the meaningless diagonal elements from the functional connectivity matrix, and then expanded the upper triangle elements of the functional connectivity matrix as the classification features. So the feature vectors dimension of functional connectivity matrix was 16 × (16 - 1) / 2 = 120.

**Feature selection**

Before applying data mining algorithm, feature selection is an important step of data processing. Removing irrelevant and redundant information often improves the performance of machine learning algorithm. When a data set has a large number of attributes, it is usually necessary to use data dimension reduction method or feature selection technology to solve "the curse-of-dimensionality"[35] caused by high-dimensional attributes. In this paper, we selected Correlation-based Feature Selection (CFS)[36, 37] with GreedyStepwise (GSW) search strategy, Information Gain and ReliefF feature selection methods to remove unimportant features.

The CFS algorithm searches feature subsets by calculating the correlation between "feature-class" and "feature-feature". The aim of this method is to find features subset that are highly correlated with the class whereas low correlated with the features, the algorithm is described as follows.

(1) Feature evaluation

The correlation between feature sets will increase with the increase of correlation coefficient between features and classes, and decrease with the increase of correlation coefficient between features. The formula is as follows:

$$\text{Merit}_s = \frac{k\overline{r_{cf}}}{\sqrt{k+k(k-1)\overline{r_{ff}}}} \tag{2}$$

Where $\text{Merit}_s$ is a heuristic index of feature subset S containing k features, called correlation coefficient, $\overline{r_{cf}}$ is the average feature-class correlation coefficient (f ∈ S), $\overline{r_{ff}}$ is the average feature-feature correlation coefficient. The numerator represents the class prediction ability of feature subset S, and the denominator represents the redundancy of features in feature subset S.

(2) Feature correlation

To calculate the correlation between features, CFS uses the measurement based on information theory[38]. If x and y are discrete random variables, formulas (3) and (4) give the entropy of Y before and after X is observed.

$$H(Y) = -\sum_{y \in Y} p(y) \log_2 p(y) \tag{3}$$

$$H(Y|X) = -\sum_{x \in X} p(x) \sum_{y \in Y} p(y|x) \log_2 p(y|x) \tag{4}$$

In the information gain, the importance of a feature is measures by calculating the difference of information entropy before and after the feature appears, if the difference is larger, the more information this feature carries, and then the more important it is. The formula of information gain is:

$$\begin{aligned} \text{gain} &= H(Y) - H(Y|X) \\ &= H(X) - H(X|Y) \\ &= H(Y) + H(X) - H(X,Y) \end{aligned} \tag{5}$$

Since the information gain is biased towards attributes with larger values, symmetric uncertainty[39] is used to compensate for this bias, and the features are normalized to between [0, 1]:

$$\text{symmetrical uncertainty} = 2.0 \times \left[\frac{gain}{H(Y)+H(X)}\right] \tag{6}$$

Therefore, CFS uses symmetric uncertainty to calculate the correlation between "feature-class" and "feature-feature" in each subset of features.

The Information Gain algorithm evaluates the worth of an attribute by measuring the information gain with respect to the class, which is calculated based on information theory as formula (3) and (4). Assume the attributes set is A and the class set is C. So Information Gain (C, A) of an attribute A relative to a collection of samples C can be calculated as:

$$\text{Information Gain }(C, A) = H(C) - H\left(\frac{C}{A}\right) \tag{7}$$

The ReliefF algorithm[40] evaluates the worth of an attribute by repeatedly sampling an instance and considering the value of the given attribute for the nearest instance of the same and different class. Assume the training set is D, the attributes set is A, the class set is C, the sampling frequency is m, and the number of nearest neighbors is k. The procedure is as below:

Input: for each training instance a vector of attribute values and the class value

Output: the feature weight W of each feature

1. Set feature weight is 0, T = φ,
2. For i := 1 to m do
   randomly select an instance $R_i$ from D;
   find k nearest neighbor of R from the same class, called nearest hits $H_j$, and find k nearest neighbors from the other class, called nearest misses $M_j(C)$;
3. For A := 1 to a do

$$W[A] := W[A] - \sum_{j=1}^{k}\frac{diff(A, R_i, H_j)}{m \cdot k} +$$

$$\sum_{C \neq class(R_i)}\left[\frac{P(C)}{1-P(class(R_i))}\sum_{j=1}^{k} diff(A, R_i, M_j(C))\right]/(m \cdot k) \tag{8}$$

End;

The Function diff(A, $R_i$, $H_j$) calculates the difference between the values of the attribute A for two instances $R_i$ and $H_j$; P(C) calculates the prior probability of class C.

**Classification**

In our experiments, we chose the classifiers including: LR, KNN, Decision Tree (DT) and Naïve Bayes (NB) to obtain classification accuracy for the EEG data with the selected features. Prior to classification, all features were normalized in the range [-1, 1] using the z–score algorithm. Due to each subject having 40 samples (40 2 epoch), the leave-one-subject-out cross-validation (LOSOCV) was used to evaluate the generalization ability of each classification model. These classifiers were set to use their default parameter values as implemented in Weka software (http://www.cs.waikato.ac.nz/ml/weka). The parameters of classifiers are shown in Table 1.

Table 1. The parameters of classifiers

| Classifier | Core parameters |
|---|---|
| LR | Ridge = 1.0E-8, maxIts = -1, numDecimalPlaces = 4 |
| KNN | k_neighbor = 3, neighborsearch = 'LinearNNsearch', distanceFunction = 'EuclideanDistance' |
| DT | MinNumObj = 2, NumFoldsPruning = 5, Seed = 1 |
| NB | Null |

## Results and discussion

### Evaluation of features set

In order to find the effective feature set for identifying depression, this paper uses four classifiers (KNN, NB, DT and LR) and three feature selection methods (CFS (GSW), Information Gain and ReliefF) to evaluate the depression recognition ability of single type feature set (linear feature (L), nonlinear feature (NL) and functional connectivity feature (PLI)) and combined type feature set (L + NL, PLI + L, PLI + NL and All (L + NL + PLI)). The results are shown in Table 2, where the values in the table represent the mean accuracy obtained by averaging the accuracy of four classifiers. It can be found that when using feature set All, the classification result is better than that of single type feature and combined type feature set L + NL. Moreover, when using PLI feature set or PLI + other types of feature set, no matter which feature selection method is used, the classification accuracy is relatively high and maintains stable. The effectiveness of PLI feature set is most obvious in LR classifier (see Fig 1a to 1d), the classification accuracy of PLI feature set is higher than that of linear feature set and nonlinear feature set. Especially when feature selection method ReliefF and classifier LR are used to classify the data containing PLI feature set, the classification accuracy can be kept at around 81% (see Fig 1d). And when the data containing All feature set is classified, the highest accuracy, sensitivity and specificity can be 82.31%, 78.14% and 90.34%, respectively.

Therefore, through these results, we can conclude that the All (L + NL +PLI) feature set is more conductive to the effective recognition of depression, which can represent more comprehensive electrophysiological characteristics of depressive patients. And we find that PLI feature set plays more important role than linear and nonlinear features in depression detection. The reasons for this phenomenon might be that the symptoms of depressed patients are related to the abnormal functional connectivity pattern among multiple brain regions, rather than to the breakdown of a single local brain region[41]. Previous studies have shown that functional connectivity analysis could reveal the abnormalities in depression[42, 43]. And the study of functional connectivity could provide significant information for the classification of depression[20, 22], so these findings further support the effectiveness of functional connectivity feature PLI in this study. In addition, in order to find a better method to identify depression, we will evaluate the performance of different classifiers and feature selection methods in classifying depression in the next section.

Table 2. Classification accuracy of different features set

|   | None | CFS (GSW) | Information Gain | ReliefF |
|---|---|---|---|---|
| L | 52.02 ± 2.99% | 67.96 ± 5.67% | 67.09 ± 7.74% | 66.88 ± 6.06% |
| **NL** | 47.31 ± 0.84% | 68.82 ± 3.40% | 68.77 ± 3.65% | 68.80 ± 5.03% |
| **L+NL** | 49.62 ± 3.57% | 68.14 ± 4.09% | 69.08 ± 5.64% | 69.73 ± 5.36% |
| **PLI** | 50.13 ± 3.96% | 70.98 ± 5.34% | 70.86 ± 6.82% | 75.08 ± 5.89% |
| **L+PLI** | 51.95 ± 4.28% | 73.24 ± 4.78% | **<u>72.64 ± 4.90%</u>** | 74.62 ± 6.64% |
| **NL+PLI** | 49.17 ± 4.50% | 72.11 ± 4.50% | 68.02 ± 5.50% | 75.20 ± 5.82% |
| **All** | **<u>52.12 ± 4.00%</u>** | **<u>73.42 ± 5.60%</u>** | 70.70 ± 5.35% | **<u>75.54 ± 6.74%</u>** |

Note: L represents linear features, NL represents nonlinear features, PLI represents phase lagging index, All represents L + NL + PLI. The value in the Table 2 represents mean accuracy of four classifiers (KNN, NB, DT and LR) classifying 24 MDD patients and 29 NC. Boldface and underline indicate the highest classification result.

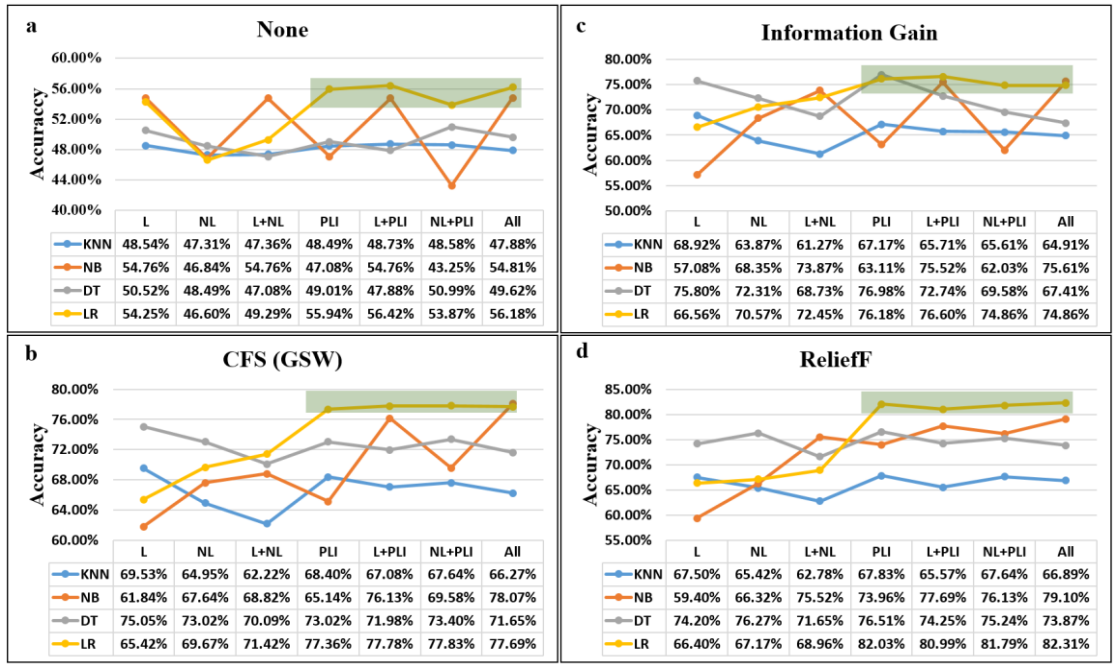

**Fig 1. Classification results of different feature set using four classifiers (KNN, NB, DT and LR) and three feature selection methods (CFS (GSW), Information Gain, ReliefF).**
Note: L represents linear features, NL represents nonlinear features, PLI represents phase lagging index, All represents L + NL + PLI.

**Evaluation of feature selection methods and classifiers**

To evaluate performance of depression recognition of different classifiers and feature selection method, we average the classification accuracy and feature length of seven kinds of feature sets used to classify 24 patients with MDD and 29 NC (see Fig 2). From the perspective of classifier, LR is the optimal classifier, followed by DT, NB and KNN (see Fig 2a). From the perspective of feature selection method, ReliefF is the optimal feature selection method, followed by CFS (GSW) and Information Gain (see Fig 2a). And the average feature length selected by ReliefF is shorter than that of the other feature selection methods (see Fig 2b). In particular, when combining classifier LR with feature selection method ReliefF to classify the data of seven kinds of feature set types, the highest average classification accuracy can be yielded 75.66%. Hence, it can be considered that LR combined with ReliefF is a preferable method to distinguish patients with MDD from NC. Moreover, both of the algorithms of LR and ReliefF are relatively simple, easy to implement, less computational, and high efficiency, so the combination method of LR and ReliefF might be very suitable for applying to the real-time system of depression detection.

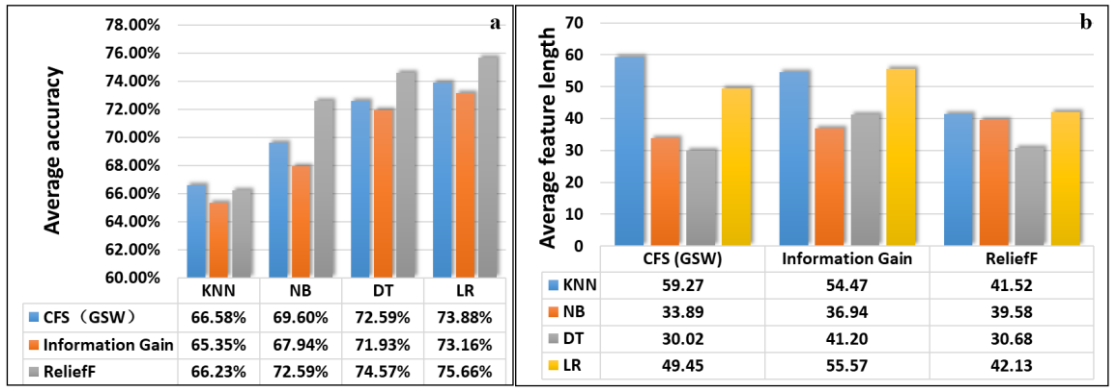

**Fig 2. The evaluation results of classifiers and feature selection methods.** (a) Average accuracy of four classifiers (KNN, NB, BT and LR): represents average of accuracy obtained by using seven feature sets (L, NL, L + NL, PLI, L + PLI, NL + PLI, All) classifying 24 MDD and 29 NC. (b) Average feature length of three feature selection methods (CFS (GSW), Information Gain and ReliefF): represents the average length of seven feature sets required by different classifiers and feature selection methods to achieve the highest classification accuracy.

**Distribution of the optimal feature set**

Table 3. Features selected from All (L + NL + PLI) feature set

| Feature type | Feature name |
|---|---|
| **Functional connectivity feature PLI (15)** | *Left hemisphere:* C3-T3, T3-P3, F3-F7, P3-T5, F7-T3, F7-C3, T3-T5, C3-P3, <br> *Right hemisphere:* P4-T6, O2-P4, O2-T6, T6-T4, P4-T4, F8-F4, C4-F8 |
| **Nonlinear feature (3)** | C0-complexity (O1, O2, T6) |

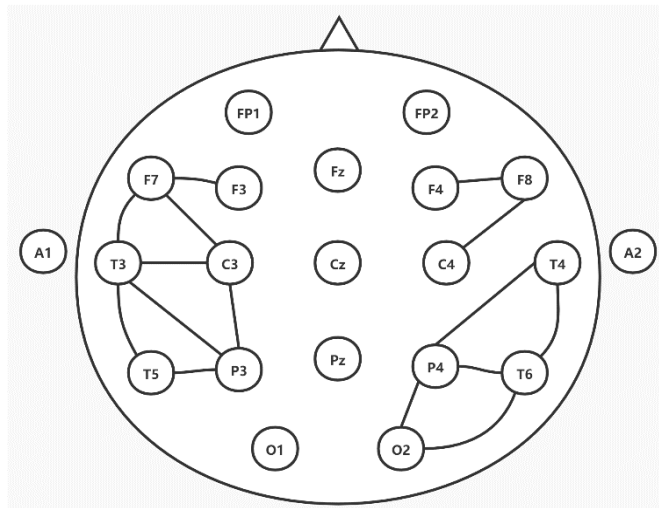

**Fig 3. Analysis of selected optimal feature distribution across the region**

According to the above results, we found that when the data containing feature set All (L + NL + PLI) is classified by using classifier LR and feature selection method ReliefF, the highest accuracy can yield 82.31%. So we further to explore the characteristic of distribution of the optimal feature set. The selected features are shown in Table 3, we can find that the number of functional connectivity features PLI is much more than the number of nonlinear and linear features, which also suggests that PLI feature set occupies an important position for classifying depression. In addition, the result of distribution of selected functional connectivity edges indicated that the intrahemispheric connection edges compared to the interhemispheric connection edges have a great effect to discriminate patients with MDD and normal controls (see Fig 3). However, whether the difference between the MDD group and the NC group does exist as mentioned above, we conducted a statistical analysis of the All feature set in the next section.

## Statistic analysis of All feature set

Table 4. The statistic result of All (L + NL + PLI) feature set between MDD group and NC group

| Feature type | Feature name |
|---|---|
| **Functional connectivity feature PLI (27)** | *Left hemisphere:* Fp1-C3, FP1-T3, FP1-P3, F3-F7, F3-C3, F3-T3, F3-P3, F7-C3, F7-T3, F7-P3, F7-T5, C3-P3, T3-T5, P3-T5, P3-O1<br>*Right hemisphere:* FP2-P4, O2-P4, O2-T6, O2-T4, P4-T4, C4-F8, C4-F4<br>*Left-Right hemisphere:* FP1-T4, T3-P4, P3-P4, T5-P4, O1-T4 |
| **Nonlinear feature (11)** | C0-complexity (C3, C4, O1, P3, P4, T4, T6), SVDen (T3); Spectral entropy (C3, O1, P4) |

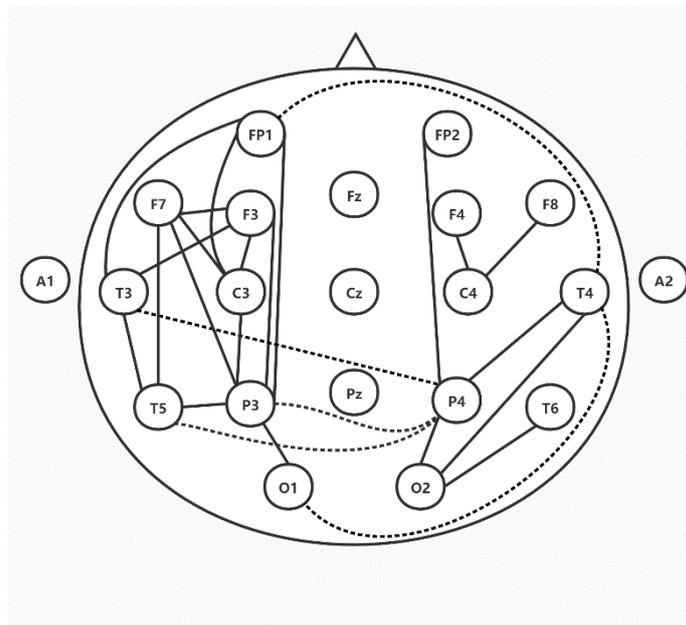

**Fig 4. The distribution of connection edges with significant difference between MDD group and NC group**

Independent sample T-test with Bonferroni correction (p < 0.05/334) was conducted to find the significant features of All (L + NL + PLI) feature set between MDD group and NC group. Statistic results are shown in Table 4. We can find that features with significant difference is still PLI features, followed by nonlinear features, and there is no significant difference in linear features. Meanwhile, we can find that having significant difference connection edges of left hemisphere are more than that of right hemisphere. There are some relevant studies have also obtained the consistent findings. For example, Henriques and Dividson et al. [44] observed that depressive patients had less left hemisphere activation than healthy subjects. Peng et al. [22] found that altered functional connectivity of depression were mainly located in the left frontal lobe and parietal lobe in the whole frequency band. Hosseinifard B et al. [10] found that there were differences in the left hemisphere of alpha frequency band between depressed patients and normal controls. And Li X et al. [45] reported that left hemispheric electrodes had significant difference to distinguish mild depressive patients from normal controls. Hence, we conclude that there exists more evident functional dysfunction in the left hemisphere of depressive patients.

Moreover, from the distribution of connection edges having significant difference between MDD group and NC group, we can also find that the intrahemispheric connection edges are much more than the interhemispheric connection edges (see Fig 4), which may indicate that there exist some differences in globing and local processing of brain between the MDD group and NC group, so future research will make further analysis from the perspective of functional brain network. All in all, these statistic results are almost consistent with the results of Table. 3 and Fig 3.

## Conclusion

With the purpose to better identify depression, this paper extracted different types of EEG features including linear features, nonlinear features and functional connectivity features PLI to comprehensively analyze the EEG signals in patients with MDD. It was found that functional connectivity feature PLI was superior to the linear features and nonlinear features. And when combining all the types of features to classify MDD patients by using the optimal combined method classifier LR with feature selection method ReliefF, we could obtain the highest classification accuracy 82.31%. We also analyzed the characteristic of distribution of the optimal feature set, results indicated that PLI did play an important role in depression recognition. Especially, intrahemispheric connection edges were much more than the interhemispheric connection edges, and intrahemispheric connection edges had a significant differences between MDD group and NC group, so intrahemispheric connection edges of PLI might be an effective biomarker to identify depression. In addition, statistics results suggested that there existed more evident functional dysfunction in the left hemisphere of depressive patients. In summary, this paper may provide reliable methods and effective biomarkers for depression detection.

## Acknowledgement

This work was supported in part by the National Key Research and Development Program of China (Grant No. 2019YFA0706200), in part by the National Natural Science Foundation of China (Grant No.61632014, No.61627808, No.61210010), in part by the National Basic Research Program of China (973 Program, Grant No.2014CB744600), and in part by the Program of Beijing Municipal Science & Technology Commission (Grant No.Z171100000117005).